\documentclass[preprint]{aastex}
\usepackage{ulem}
\usepackage{bm}
\usepackage{amsmath}

\shorttitle{Radial evolution of solar wind turbulence anisotropy} \shortauthors{He et al.}

\begin{document}

\title{Radial evolution of the wave-vector anisotropy of solar wind turbulence between 0.3 and 1~AU}

\author{Jiansen He\altaffilmark{1,2}, Chuanyi Tu\altaffilmark{1},
Eckart Marsch\altaffilmark{3}, Sofiane Bourouaine\altaffilmark{4}, Zhongtian
Pei\altaffilmark{1}}

\altaffiltext{1}{Department of Geophysics, Peking University, Beijing,
100871, China; E-mail: jshept@gmail.com} \altaffiltext{2}{State Key
Laboratory of Space Weather, Chinese Academy of Sciences, Beijing 100190}
\altaffiltext{3}{Institute for Experimental and Applied Physics, Christian
Albrechts University at Kiel, 24118 Kiel, Germany} \altaffiltext{4}{Space
Science Center and Department of Physics, University of New Hampshire,
Durham, NH 03824, USA}

\begin{abstract}
We present observations of the power spectral anisotropy in wave-vector space of
solar wind turbulence, and study how it evolves in interplanetary space with
increasing heliocentric distance. For this purpose we use magnetic field
measurements made by the Helios-2 spacecraft at three positions between 0.29 and
0.9~AU. To derive the power spectral density (PSD) in $(k_\parallel,
k_\bot)$-space based on single-satellite measurements is a challenging
task not yet accomplished previously. Here we derive the spectrum
$\rm{PSD}_{\rm{2D}}$($\rm{k}_\parallel$, $\rm{k}_\bot$) from the spatial
correlation function $\rm{CF}_{\rm{2D}}(r_\parallel, r_\bot)$ by a
transformation according to the projection-slice theorem. We find the so constructed PSDs to be
distributed in k-space mainly along a ridge that is more inclined toward the
$\rm{k}_\bot$ than $\rm{k}_\parallel$ axis, a new result which probably
indicates preferential cascading of turbulent energy along the $\rm{k}_\bot$
direction. Furthermore, this ridge of the distribution is found to gradually
get closer to the $\rm{k}_\bot$ axis, as the outer scale
length of the turbulence becomes larger while the solar wind flows further
away from the Sun. In the vicinity of the $\rm{k}_\parallel$ axis, there
appears a minor spectral component that probably corresponds to
quasi-parallel Alfv\'enic fluctuations. Their relative contribution to the
total spectral density tends to decrease with radial distance. These findings
suggest that solar wind turbulence undergoes an anisotropic cascade
transporting most of its magnetic energy towards larger $\rm{k}_\bot$, and
that the anisotropy in the inertial range is radially developing further at
scales that are relatively far from the ever increasing outer scale.
\end{abstract}


\section{Introduction}

Solar wind fluctuations are considered as the genuine and prominent example
of magnetohydrodynamic (MHD) turbulence \citep[e.g.,][]{Tu1995SSRv,
Goldstein1995ARA&A, Bruno2005LRSP, Marsch2006LRSP}, which is driven by solar
activity and naturally occurs in the inhomogeneous interplanetary space.
Through in situ measurements made by the Helios 1 and Helios 2 spacecraft,
the space plasma physics community has gained abundant knowledge about the
radial evolution of solar wind turbulence in the inner heliosphere. Magnetic
power spectra of fluctuations at MHD scales were found
\citep{Bavassano1982JGR} to show two separate frequency ranges, whereby the
power in the higher-frequency ($\rm{f}>2.5\times10^{-3}$~Hz in the spacecraft
frame) range decays with radial distance (like $\rm{r}^{-4.2}$) faster than
that in the lower-frequency ($\rm{f}<2.5\times10^{-3}$) range (with radial
scaling like $\rm{r}^{-3.2}$).

The radial evolution of the lower-frequency magnetic power spectra can be
reproduced by the WKB-theory of Alfv\'en wave propagation
\citep{Whang1973JGR, Hollweg1974JGR}, which predicts a similar radial
evolution. Whereas the higher-frequency magnetic power spectra, which show a
steeper profile (Kolmogrov-like) with its spectral break frequency shifting
towards lower values during the radial evolution, were successfully
reproduced by Tu's turbulence model \citep{Tu1984JGR, Tu1988JGR}, which took
into account (together with the WKB description) the nonlinear interaction
between counter-propagating imbalanced Alfv\'en waves. Moreover, the
normalized cross-helicity (Alfv\'enicity) was shown to decrease with
increasing heliocentric distance \citep{Roberts1987JGR,
Marsch1990JGR,Grappin1990JGR}, which to explain was beyond the scope of Tu's
model. To self-consistently describe the radial evolution of turbulent
energy, cross-helicity, and Alfv\'en ratio, substantial theoretical efforts
had to be made, which finally resulted in general transport equations
\citep{Marsch1989JPlPh, Tu1990JPlPh, Zhou1989GeoRL} for the related spectra.

In numerical simulations of MHD turbulence, the assumed background magnetic
field ($\mathbf{B}_0$) was found to cause spatial anisotropy of the turbulent
fluctuations along and across the mean field, with the parallel scale
generally being larger than the perpendicular scale
\citep[e.g.,][]{Shebalin1983JPlPh, Biskamp2000PhPl, Cho2002ApJ}. For balanced
strong MHD turbulence with vanishing cross-helicity, the anisotropy is
predicted to reveal a scaling relation obeying
$\rm{k}_\parallel\sim\rm{k}_\bot^{2/3}$, which was derived in a
phenomenological theory \citep{Goldreich1995ApJ} based on the conjecture of
critical balance, i.e. the rough equality between the linear wave-propagation
time and nonlinear eddy-interaction time. Numerical simulations further
showed that balanced strong turbulence behaves for strong or weak
$\mathbf{B}_0$ differently in its scaling properties across $\mathbf{B}_0$:
Iroshnikov-Kraichnan scaling was found for strong $\mathbf{B}_0$ and
Goldreich-Shridar scaling for weak $\mathbf{B}_0$ \citep{Mueller2003PhRvE}.
These different scalings are argued to be probably attributed to an increase
of dynamic alignment as the cascade proceeds to smaller scale, which may also
induce scaling anisotropy in the plane perpendicular to $\mathbf{B}_0$
\citep{Boldyrev2005ApJ}.

However, in the solar wind and particular in fast streams one usually
observes imbalanced turbulence with outgoing waves dominating over incoming
waves. This imbalanced turbulence implies different nonlinear interaction
time scales for the oppositely propagating waves, and is more complex than
the balanced one. Its physical nature remains a controversial issue, although
several theories have been proposed \citep{Lithwick2007ApJ, Beresnyak2008ApJ,
Chandran2008ApJ, Podesta2010ApJ}.

The spatial anisotropy of solar wind turbulence was studied by means of data
analysis employing various tools, such as correlation function
\citep{Matthaeus1990JGR, Dasso2005ApJ, Osman2007ApJ, Nemeth2010SoPh},
structure function \citep{Luo2010ApJ, Chen2010PhRvL, Chen2012ApJ}, power of
magnetic components \citep{Bieber1996JGR}, and power law scaling
\citep{Horbury2008PhRvL, Podesta2009ApJ, Wicks2010MNRAS, Wicks2011PhRvL}.
Scaling anisotropy becomes more clearly visible if one uses a scale-dependent
local mean magnetic field ($\mathbf{B}_{0,\rm{local}}$) (which was first
introduced by \citet{Horbury2008PhRvL} applying the wavelet technique)
instead of a constant global mean magnetic field \citep{Tessein2009ApJ}. Some
efforts have been also made to reconstruct magnetic PSD in multi-dimensional
wave-vector space by means of the k-filtering method
\citep{Sahraoui2010PhRvL, Narita2010PhRvL}, which was developed originally to
distinguish a limited number of plane waves from multi-position measurements
\citep{Pincon1991JGR}. The integrated $\rm{PSD}_{\rm{1D}}(\rm{k}_\bot)$ with
a spectral index $\sim-1.6$ as obtained from k-filtering method seems to
resemble the reduced $\rm{PSD}_{\rm{1D}}(\rm{k}_\bot)$ as derived from the
direct wavelet transformation. However, in some cases studied with the
k-filtering method, the integrated $\rm{PSD}_{\rm{1D}}(\rm{k}_\parallel)$
shows a spectral index $\sim-7.0$ \citep{Sahraoui2010PhRvL}, differing
significantly from the reduced $\rm{PSD}_{\rm{1D}}(\rm{k}_\parallel)$
($\sim\rm{k}_\parallel^{-2}$) as obtained by the wavelet method
\citep{Horbury2008PhRvL}. The reliability of k-filtering method for
estimating turbulent power spectra may need further validation, e.g. by
applying it to numerically simulated turbulence with known scaling.

Previous studies have revealed the evolution of reduced 1D-PSD in the inner
heliosphere, and have presented evidence of wave-vector anisotropy at
specific positions [e.g., 1~AU]. However, the turbulence anisotropy pattern
at 0.3~AU (the innermost distance reached in-situ so far) and its evolution
trend between 0.3 and 1.0~AU has not yet been investigated. To do this is an
important task, because it will provide the needed information about the
evolution of the energy cascading route in k-space, and reveal possible ways
of turbulent energy dissipation required for sustained solar wind heating.
This work is dedicated to a study of MHD turbulence anisotropy and will
provide new knowledge on its spectral characteristics. The data analysis to
achieve these goals is briefly described as follows.

Firstly, we estimate the second-order structure function as a function of
$\theta_{\rm{RB}}$ (the angle between the radial direction and the local mean
magnetic field vector). Accordingly, the angular distribution of the spatial
correlation function is obtained, using the relation between structure
function and correlation function. Secondly, we fit the measured structure
function with a compound fit function, resembling a power-law dependence at
short scale and giving an exponential trend at large scale. The fitted
angular correlation function is subsequently derived. Thirdly, under the
assumption of a statistically time-stationary state, the relative 2D-PSD in
$(\rm{k}_\parallel, \rm{k}_\bot)$ space is constructed from the fitted
angular distribution of the correlation function, whereby we make use of the
projection-slice theorem which is fundamental for image processing in medical
tomography (see \citet{Bovik2000} for a detailed review).

\section{Analysis method}

In this section, we describe the applied methods and the data analysis, which
includes: how to derive angular distributions of the structure function
$\rm{SF}(\tau, \theta_{\rm{VB}})$ and the correlation function $\rm{CF}(\tau,
\theta_{\rm{VB}})$; how to fit $\rm{SF}(\tau, \theta_{\rm{VB}})$ and
$\rm{CF}(\tau, \theta_{\rm{VB}})$ appropriately; and how to obtain
$\rm{PSD}_{\rm{2D}}(k_\parallel, k_\bot)$ as transformed from $\rm{CF}(\tau,
\theta_{\rm{VB}})$, which is in turn obtained from $\rm{CF}(r_\parallel,
r_\bot)$ by assuming a quasi-steady state with $r_\parallel \simeq
V_{\rm{sw}}\tau\cos\theta_{\rm{VB}}$ and $r_\bot \simeq V_{\rm{sw}}\tau
\sin\theta_{\rm{VB}}$, corresponding to the Taylor assumption of fluctuations
being frozen into the flow, and thus being simply convected by the wind past
the spacecraft. Here $V_{\rm{sw}}$ is the solar wind speed.

The second-order magnetic structure function is defined as the ensemble
average of the squared magnetic field vector difference. It can be written as
\begin{equation}
{\rm{SF}}(\tau ) = \left\langle {( {\mathbf{B}(t + \frac{\tau }{2})
- \mathbf{B}(t - \frac{\tau }{2})} )^2 } \right\rangle ,\label{Eq.1}
\end{equation}
where the angular bracket denotes in practice a time average in our
subsequent data analysis. This time average permits one to quantify the
global scaling of the magnetic fluctuations, without distinguishing a
possible scaling-law difference for different angles ($\theta_{\rm{VB}}$)
between the sampling direction and the local mean magnetic field vector
($\mathbf{B}_{0,\rm{local}}$). The local mean magnetic field is known to be
changing in time and depend on scale ($\mathbf{B}_{0,\rm{local}}(t, \tau)$),
leading to a scale-dependent variation of the angle $\theta_{\rm{VB}}$ with
time. For solar wind with a radial speed much larger than the velocity
fluctuation amplitude, the quantity $\theta_{\rm{VB}}$ can be approximated by
$\theta_{\rm{RB}}$ (i.e., the angle between the radial direction and the
$\mathbf{B}_{0,\rm{local}}$ direction), which is used hereafter. To estimate
the structure function value at a certain time scale $\tau'$ and for a
certain angle $\theta'_{\rm{RB}}$, one needs to pick out the values of
${\rm{SF}}(t,\tau')$ at those times when
$\theta_{\rm{RB}}(t,\tau')=\theta'_{\rm{RB}}$, and then make an average over
all the so picked samples. Therefore, the corresponding angular distribution
of the structure function can be expressed as
\begin{equation}
{\rm{SF}}(\tau ',\theta '_{\rm{RB}} ) =
\frac{ \int_0^T {(\mathbf{B}(t + \frac{{\tau '}}{2}) - \mathbf{B}(t - \frac{{\tau '}}{2}))^2 dt
\left| {_{\theta _{\rm{RB}}(t,\tau') = \theta '_{\rm{RB}} } } \right.} }{ \int_0^T {dt \left| {_{\theta _{\rm{RB}}(t,\tau')
= \theta '_{\rm{RB}} } } \right.} }.\label{Eq.2}
\end{equation}
Here the time period for the whole chosen data set is indicated as $T$. It
should be much larger than the time scale $\tau^\prime$, and thus we may
formally take the limit $T \rightarrow \infty$.

Expressing the ensemble average used in equation (\ref{Eq.1}) explicitly as a
time average, the relation between the structure function $\rm{SF}(\tau)$ and
the correlation function $\rm{CF}(\tau)$ can be obtained from the subsequent
calculation:
\begin{equation}
\label{Eq.3}
\begin{split}
  {\rm{SF}}(\tau ) &= \frac{1}{T} \int_0^T (\mathbf{B}(t + \frac{\tau }{2}) - \mathbf{B}(t - \frac{\tau }{2}))^2 dt  \\
  &= \frac{1}{T} \left[ \int_0^T \mathbf{B}^2 (t + \frac{\tau }{2})dt  + \int_0^T \mathbf{B}^2 (t - \frac{\tau }{2})dt  - 2\int_0^T \mathbf{B}(t + \frac{\tau }{2})\cdot\mathbf{B}(t - \frac{\tau }{2})dt \right] \\
  &= 2\rm{CF}(\tau = 0) - 2\rm{CF}(\tau ). \\
\end{split}
\end{equation}
Using the above definition (\ref{Eq.2}), the angular distribution of the
correlation function can also be approximated by the angular distribution of
the structure function, yielding on the basis of (\ref{Eq.3}) the following
relation:
\begin{equation}
\label{Eq.4}
\begin{split}
  \rm{SF}(\tau ,\theta ') &=  - 2\left\langle \mathbf{B}(t + \frac{\tau }{2}) \cdot \mathbf{B}(t - \frac{\tau }{2}) \right\rangle \left| {_{\theta _{RB}  = \theta '} } \right. + \left\langle \mathbf{B}^2 (t + \frac{\tau }{2}) \right\rangle \left| {_{\theta _{RB}  = \theta '} } \right. + \left\langle \mathbf{B}^2 (t - \frac{\tau }{2}) \right\rangle \left| {_{\theta _{RB}  = \theta '} } \right. \\
  &\simeq  - 2 \,\rm{CF}(\tau ,\theta ') + 2 \,\rm{CF}(\tau  = 0, \theta ') \\
  &\simeq  - 2 \,\rm{CF}(\tau ,\theta ') + 2 \,\rm{CF}(\tau  = 0), \\
\end{split}
\end{equation}
where angular isotropy of $\rm{CF}$ at $\tau=0$ was assumed in the
derivation. Under Taylor's hypothesis the solar wind fluctuations can be
considered time stationary, as the wave phase speed is small in comparison to
the supersonic convection speed, and then $\rm{CF}(\tau, \theta')$ can be
rewritten as a spatial correlation function in the 2D r-space,
\begin{equation}
\label{Eq.5}
\begin{split}
\rm{CF}(\tau ,\theta ') \sim \rm{CF}_{\rm{2D}} (r_\parallel  ,r_ \bot  ),
\end{split}
\end{equation}
with $r_\parallel=V_{\rm{sw}}\tau \cos\theta'$ and $r_\bot=V_{\rm{sw}}\tau
\sin\theta'$. This completes the derivation of the two-dimensional
correlation function from the structure function. We note that the frozen-in-flow Taylor's hypothesis may be slightly weakened for smaller heliocentric distance with smaller Alfv\'en Mach number, which drops from higher than 10 at 1~AU to 3-4 near 0.29~AU. The quantity of main
interest is the power spectral density
$\rm{PSD}_{\rm{2D}}(k_\parallel,k_\bot)$, which in can in principle be
obtained directly from Fourier transformation of
$\rm{CF}_{\rm{2D}}(r_\parallel, r_\bot)$ as follows:
\begin{equation}
\rm{PSD}_{\rm{2D}} (k_\parallel  ,k_ \bot  ) = \int_{ - \infty }^{ + \infty }{\int_{ - \infty }^{ + \infty } {\rm{CF}_{\rm{2D}} (r_\parallel  ,r_ \bot  )\exp ( - i(k_\parallel  r_\parallel   + k_ \bot  r_ \bot  ))dr_\parallel  dr_ \bot  } }.\label{Eq.6}
\end{equation}
However, we take here a new route to estimate
$\rm{PSD}_{\rm{2D}}(k_\parallel,k_\bot)$. It can also be derived from the
projected (integrated) 1D correlation function on the basis of the
projection-slice theorem \citep{Bovik2000} with help of the following
formula:
\begin{equation}
\label{Eq.7}
\begin{split}
\rm{PSD}_{\rm{2D}} (k,\theta_k ) &= \int_{ - \infty }^{ + \infty } {\int_{ - \infty }^{ + \infty } {\rm{CF}_{\rm{2D}} (r_\parallel  ,r_ \bot  )\exp ( - i(k(r_\parallel  \cos \theta _k  + r_ \bot  \sin \theta _k )))dr_\parallel  dr_ \bot  } }  \\
  &= \int_{ - \infty }^{ + \infty } {\int_{ - \infty }^{ + \infty } {\rm{CF}_{\rm{2D}} (r'\cos \theta _k  - u'\sin \theta _k ,r'\sin \theta _k  + u'\cos \theta _k )\exp ( - i(kr'))dr'du'} }  \\
  &= \int_{ - \infty }^{ + \infty } {\rm{CF}_{\rm{1D}} (r';\theta _k )\exp ( - i(kr'))dr'},  \\
\end{split}
\end{equation}
where $\theta_k$ is the angle between $\mathbf{k}$ and
$\mathbf{B}_{0,\rm{local}}$, and $\rm{CF}_{\rm{1D}}(r';\theta_k)$ is the 1D
projection (integration) of $\rm{CF}_{\rm{2D}}(r_\parallel,r_\bot)$ along the
direction normal to $\mathbf{k}$,
\begin{equation}
\rm{CF}_{\rm{1D}} (r';\theta _k ) = \int_{ - \infty }^{ + \infty } {\rm{CF}_{\rm{2D}} (r'\cos \theta _k  - u'\sin \theta _k ,r'\sin \theta _k  - u'\cos \theta _k )du'}. \label{Eq.8}
\end{equation}

Therefore, there are two approaches to calculate
$\rm{PSD}_{\rm{2D}}(k_\parallel, k_\bot)$, one may adopt either
Equation~\ref{Eq.6} or \ref{Eq.7}. In practice, the estimation of the 2D
correlation function with help of Equation~\ref{Eq.6} introduces some
uncertainty, as the noise involved in the data may destroy the required
positivity of the $\rm{PSD}$ in the entire $(k_\parallel, k_\bot)$ space. To
guarantee this positivity of $\rm{PSD}$ everywhere, one needs to approximate
the $\rm{CF}$ with some kind of positive-definite fit function before the
Fourier transformation. It is hard to find an adequate function that globally
fits the observed $\rm{CF}_{\rm{2D}}(r_\parallel,r_\bot)$ well, whereas it is
relatively easy to choose a proper fitting function for the projected
$\rm{CF}_{\rm{1D}}(r; \theta_k)$. Therefore, in our work, we will use a
fitted $\rm{CF}_{\rm{1D}}(r; \theta_k)$ to reconstruct reliably the
$\rm{PSD}_{\rm{2D}}(k_\parallel, k_\bot)$ according to Equation~\ref{Eq.7}.

\begin{figure}
\centering
\includegraphics[width=14cm]{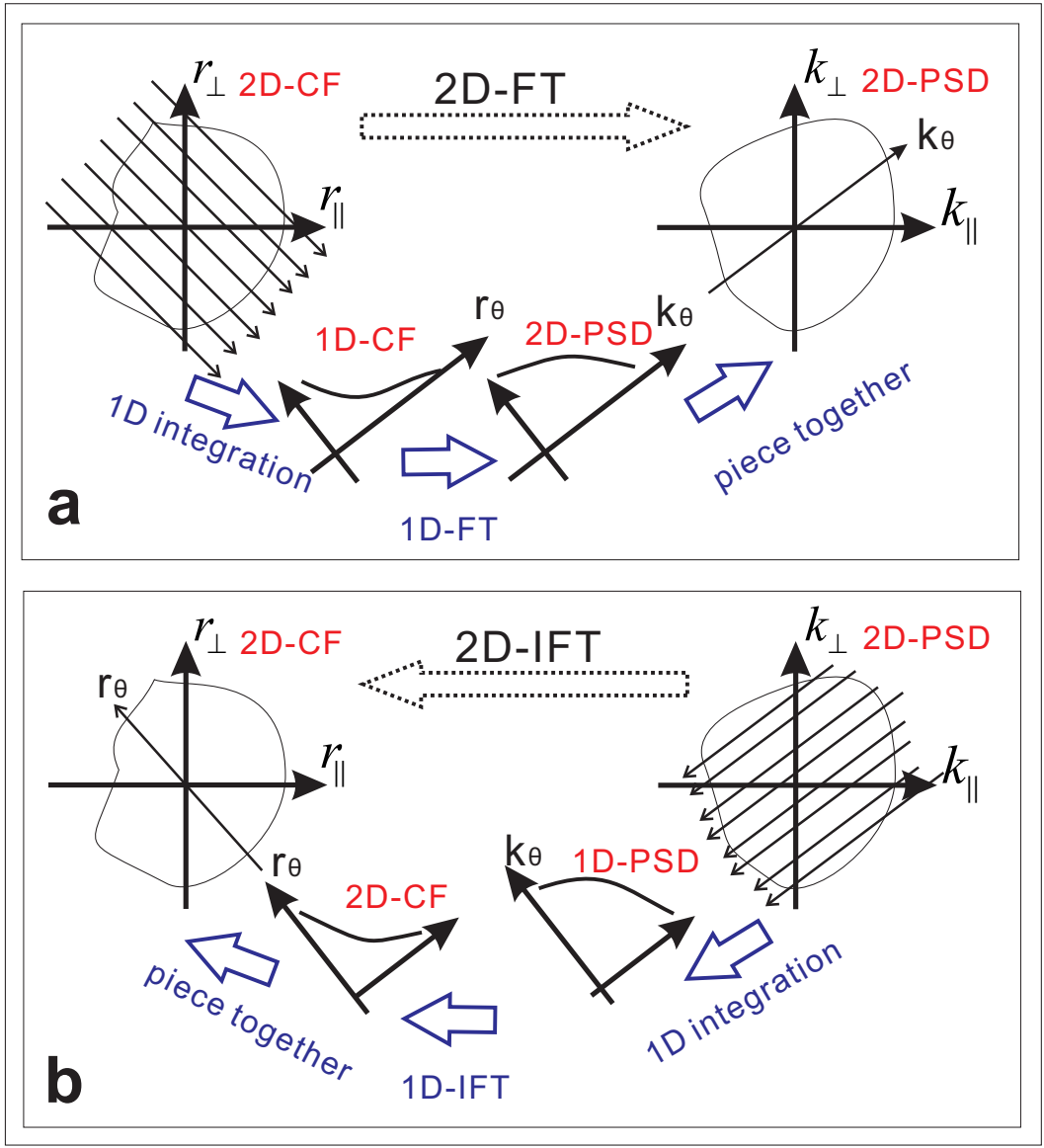}
\caption{Sketch of relation between 2D-PSD and 2D-CF based on the projection-slice
theorem. (Top) Two approaches to derive $\rm{PSD}_{\rm{2D}}(k_\parallel,k_\bot)$
from $\rm{CF}_{\rm{2D}}(r_\parallel, r_\bot)$: direct 2D Fourier transform and
indirect Fourier transform of the projected $\rm{CF}_{\rm{1D}}(r, \theta)$.
(Bottom) Vice versa for the derivation of $\rm{CF}_{\rm{2D}}(r_\parallel,r_\bot)$
from $\rm{PSD}_{\rm{2D}}(k_\parallel,k_\bot)$}
\label{Fig.1}
\end{figure}

To provide the reader with an intuitive impression about the relations
between $\rm{CF}_{\rm{2D}}(r_\parallel, r_\bot)$, $\rm{CF}_{\rm{1D}}(r,
\theta_k)$, and $\rm{PSD}_{\rm{2D}}(k_\parallel, k_\bot)$, we present the
schematic illustration shown in the upper panel of Figure~\ref{Fig.1}, which
explains the two roads from $\rm{CF}_{\rm{2D}}$ to $\rm{PSD}_{\rm{2D}}$
(direct 2D Fourier transform and indirect method based on the
projection-slice theorem). Similarly, one slice of $\rm{CF}_{\rm{2D}}$ at
certain angle $\theta_r$ is also the 1D inverse Fourier transform of
$\rm{PSD}_{\rm{1D}}$ as projected from $\rm{PSD}_{\rm{2D}}$ onto the
corresponding direction $\mathbf{k}$ with $\theta_{\rm{kB}}=\theta_r$, an
approach which is displayed in the lower panel of Figure~\ref{Fig.1}. The
relation between $\rm{CF}_{\rm{2D}}$ and $\rm{PSD}_{\rm{1D}}$ is the basic
method for calculating $\rm{CF}_{\rm{2D}}$, which was used in previous
studies \citep{Matthaeus1990JGR, Dasso2005ApJ, Osman2007ApJ}. In principle, it is also possible to derive $\rm{PSD}_{\rm{2D}}$ from $\rm{PSD}_{\rm{1D}}$ according to the method of inverse Radon transform (filtered back-projection) (private communication with M. Forman). However, this method fails in a typical benchmark test due to extreme large PSD at small $|\mathbf{k}|$, which blurs the entire reconstructed $\rm{PSD}_{\rm{2D}}$ thereby destroying its original pattern.

Speaking of the fitting function for the $\rm{CF}$, we need to mention also
the fitting function for the $\rm{SF}$, which is used to reproduce the key
features of the $\rm{SF}$. For example, people usually adopt an exponential
function to fit the profile of the $\rm{SF}$ at large scale, while they use a
power-law function for the small-scale trend. However, as far as we know,
there exists no attempt to describe both the small-scale power-law trend and
the large-scale exponential trend simultaneously with a single fitting
function. To fulfill this task, we suggest a compound function,
\begin{equation}
\rm{SF}(\tau) = 2R_0  \cdot [1 - \exp ( - (\frac{\tau }{{\tau _c }})^p )],\label{Eq.9}
\end{equation}
which interpolates between these limits. There are three parameters to be
fitted: $R_0$ means the auto-covariance at $\tau=0$, $\tau_c$ represents the
correlation time at large scale, and the index $p$ describes the power-law
scaling at short scale. Generally, for $\rm{SF}(\tau,\theta_{\rm{RB}})$ at
different $\theta_{\rm{RB}}$, the parameters $R_0$ and $\tau_c$ do not change
a lot, while $p$ remains variable. Therefore, in our practice, $R_0$ and
$\tau_c$ are obtained by fitting the time-averaged $\rm{SF}(\tau)$, and then
$p$ is determined at various $\theta_{\rm{RB}}$ by fitting $\rm{SF}(\tau,
\theta_{\rm{RB}})$, but only after $R_0$ and $\tau_c$ were set. Another
practical reason for presetting $R_0$ and $\tau_c$ before fitting
$\rm{SF}(\tau, \theta_{\rm{RB}})$ is that for every $\theta_{\rm{RB}}$ the
calculated $\rm{SF}(\tau, \theta_{\rm{RB}})$ is usually unable to reach to
the outer scale.

\section{Data analysis results}

The magnetic data (with a time resolution of about 0.25~s) used here is from
measurements by Helios-2 spacecraft at three radial positions (0.29, 0.65,
and 0.87~AU) during three time intervals (day of year: 106-109, 76-78, and
49-51 in 1976). The solar wind streams explored during these time intervals
are known to be recurrent streams emanating from a common source region on
the Sun \citep{Bavassano1982JGR}. The corresponding radial evolution of 1D
reduced magnetic PSD was presented in that paper, which observationally
promoted the development of the WKB-like solar wind turbulence model
\citep{Tu1984JGR}. Three decades later, we analyse the same data set again,
but for the purpose of revealing the evolution of solar wind turbulence in
terms of its wave-vector anisotropy.

\begin{figure}
\centering
\includegraphics[width=17cm]{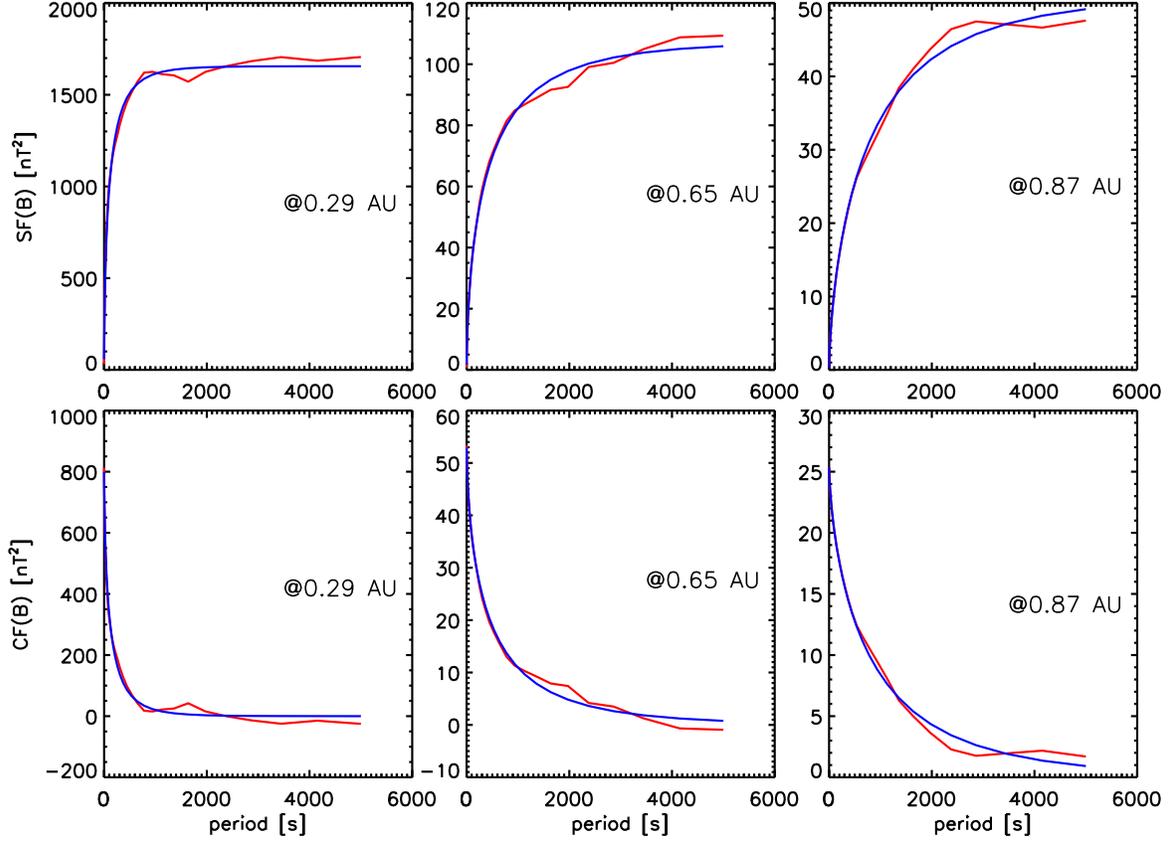}
\caption{(Top) Time averaged second-order structure functions based on
Equation~\ref{Eq.1} (red) at three positions (0.29, 0.65, and 0.87~AU) and
their corresponding fitting results according to Equation~\ref{Eq.9} (blue).
The sets of the three fitting parameters ($R_0~[\rm{nT}^2]$, $t_c~[s]$, and $p$)
are (827, 116, 0.61), (53, 465, 0.61), and (25, 857, 0.67) at 0.29, 0.65, and 0.87~AU.
(Bottom) Corresponding correlation functions based on Equation~\ref{Eq.3}
(estimations in red, fitting results in blue).}
\label{Fig.2}
\end{figure}

We use Equation~\ref{Eq.2} to estimate the second-order structure function
$\rm{SF}(\tau)$. It is defined as the magnetic vector difference squared,
which is averaged respectively over the three time intervals of our data set.
During the estimation, the data gaps are excluded without making any type of interpolation. The top three panels of Figure~\ref{Fig.2} illustrate the estimation results
as red curves. The blue lines are fitting results based on
Equation~\ref{Eq.9}, which basically match the estimates at both small and
large scales. The fitting parameters ($R_0$[$\rm{nT}^2$], $t_c$[s], $p$) at
three radial positions are found to be: (827, 116, 0.61), (53, 465, 0.61),
and (25, 857, 0.67), respectively. The fitting parameter $t_c$ (corresponding
to the correlation time) increases with heliographic distance. The values of
the exponent $p$ relate to the power-law index ($\sim-(p+1)$) of the
corresponding PSDs, which is found to be around $-1.6$, i.e. near the
Kolmogorov value of $-5/3$. The bottom three panels of Figure~\ref{Fig.2}
show the corresponding correlation function $\rm{CF}(\tau)$ as derived from
Equation~\ref{Eq.3}.

\begin{figure}
\centering
\includegraphics[width=17cm]{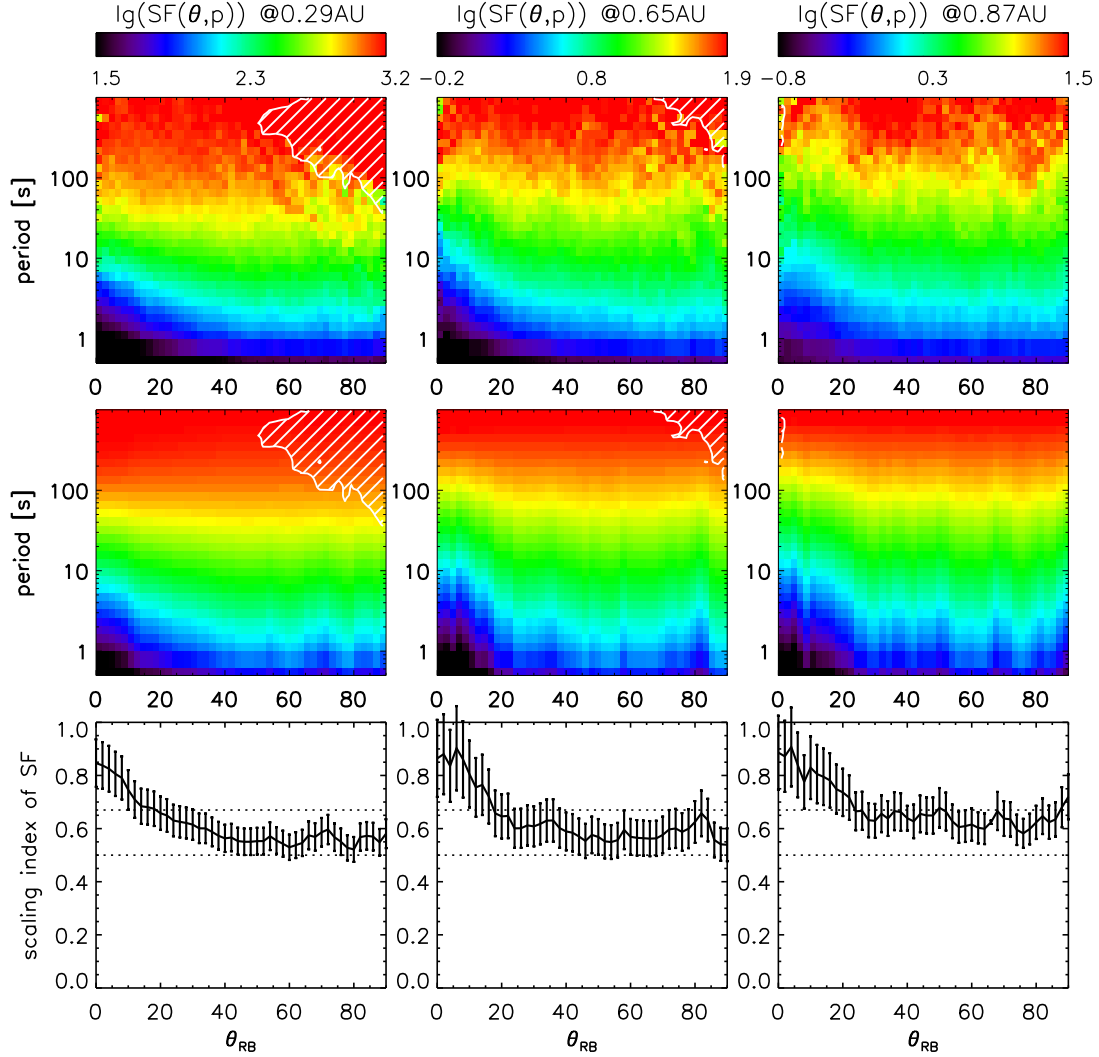}
\caption{(Top) Angular distribution of second-order structure functions $\rm{SF}(\tau,\theta_{\rm{RB}})$
estimated on the basis of Equation~\ref{Eq.2}. (Middle) Fitting results
for $\rm{SF}(\tau,\theta_{\rm{RB}})$ with $\rm{SF}(\tau)$ at every $\theta_{\rm{RB}}$
being fitted according to Equation~\ref{Eq.9}, whereby $p$ is fitted in angular dependence.
(Bottom) Fit parameter $p$ as a function of $\theta_{\rm{RB}}$, revealing
the scaling anisotropy of the structure function. The error-bars denote the fitting errors of the parameter $p$.}
\label{Fig.3}
\end{figure}

We calculate the structure functions in the angular dimension as a function
of $\theta_{\rm{RB}}$ according to Equation~\ref{Eq.2}, and display them in
the first row of Figure~\ref{Fig.3}. Apparently, the distribution of
$\rm{SF}$ is not uniform in the angle range between $0^\circ$ and $90^\circ$,
with a lower level near $0^\circ$. The non-uniform angular distribution is
more significant at short scales [e.g., $<100$~s]. For $\rm{SF}(\tau,
\theta_{\rm{RB}})$ at larger scales ($\tau>100$~s), it gradually changes from
uniformity at 0.29~AU to non-uniformity at 0.87~AU. This angular
non-uniformity is a feature hinting at anisotropy of the power spectrum in
the wave-vector space. Likewise, the extension of the angular non-uniformity
towards larger scales indicates that the wave-vector anisotropy of
larger-scale fluctuations evolves as heliocentric distance increases.

We also fit the estimated structure function by the function $\rm{SF}(\tau,
\theta_{\rm{RB}})$ of Equation~\ref{Eq.9}. To make sure the fitting process
converges for every angle, we restrict the number of fitting parameters to
$p$, while we fix the other two parameters ($R_0$ and $t_c$), both of which
may be regarded as constant without angular dependence. The fitted angular
distributions are illustrated in the second row of Figure~\ref{Fig.3}, which
look similar to the observations. The angular dependence of the fit parameter
$p$ is plotted in the third row, showing that the angular variation of
$\rm{SF}(\tau, \theta_{\rm{RB}})$ is non-uniform not only in magnitude (first
row in Figure~\ref{Fig.3}) but also in the scaling index (third row). We note
that $\rm{SF}(\tau, \theta_{\rm{RB}})$ as shown in Figure~\ref{Fig.3} relates
to the squared module of the magnetic-vector difference ($\delta B^2_x +
\delta B^2_y + \delta B^2_z$). The structure function $\rm{SF}(\tau,
\theta_{\rm{RB}})$ for the component $\delta B^2_\parallel$ (parallel to
$\mathbf{B}_{\rm{0,local}}$) shows a similar non-uniform angular dependence.
However, the calculated $\rm{SF}$ for $\delta B^2_\parallel$ has a plain
segment starting at small $\tau$, and cannot be fitted well by the function
of Equation~\ref{Eq.9}.

\begin{figure}
\centering
\includegraphics[width=17cm]{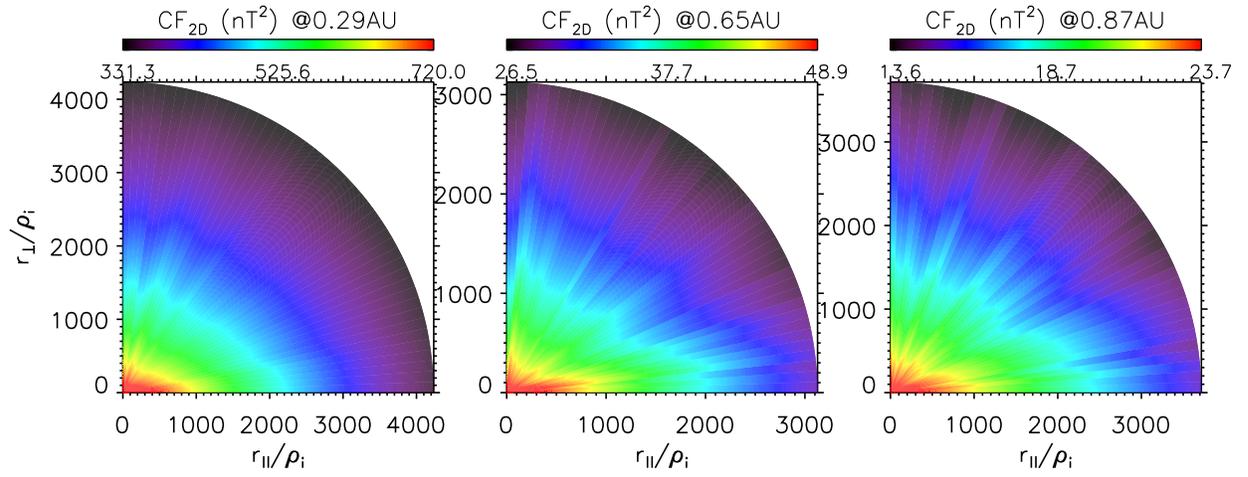}
\caption{Angular distribution of $\rm{CF}_{\rm{2D}}$ as derived from $\rm{SF}(\tau,\theta_{\rm{RB}})$
according to Equation~\ref{Eq.4} and displayed in $(r_\parallel, r_\bot)$ space
under the Taylor hypothesis of near time-stationarity. An elongation of $\rm{CF}$
along $r_\parallel$, implying the location of most turbulent energy close to $k_\bot$, is
revealed at all positions. The proton gyroradius $\rho_{\rm{g}}$ is 17, 48,
and 70~km at 0.29, 0.65, and 0.87~AU. It is used for normalization of the
spatial coordinates.}
\label{Fig.4}
\end{figure}

The angular distribution of the correlation function $\rm{CF}(\tau,
\theta_{\rm{RB}})$ is derived from the fit function $\rm{SF}(\tau,
\theta_{\rm{RB}})$ according to Equation~\ref{Eq.4}. In the light of the
projection-slice theorem as applied to the relationship between the 2D
functions $\rm{CF}$ and $\rm{PSD}$ (lower panel in Figure~\ref{Fig.1}), the
quantity $\rm{CF}(\tau, \theta_{\rm{RB}})$ is essentially a 2D correlation
function $\rm{CF}(r_\parallel, r_\bot)$, which is in principle an inverse
Fourier transform of the 2D $\rm{PSD}(k_\parallel, k_\bot)$ yet not known. In
Figure~\ref{Fig.4}, we plot the resulting $\rm{CF}(r_\parallel, r_\bot)$. The
coordinates of the abscissa ($r_\parallel$) and ordinate ($r_\bot$) are
estimated by $r_\parallel=V_{\rm{sw}}\tau\cos\theta_{\rm{RB}}$ and
$r_\bot=V_{\rm{sw}}\tau\sin\theta_{\rm{RB}}$, respectively. The main part of
$\rm{CF}(r_\parallel,r_\bot)$ is elongated along $r_\parallel$, which is
similar to the ``2D'' population of the so-called Maltese cross
\citep{Matthaeus1990JGR}. However, the ``slab'' population, which was
reported in previous statistical studies of $\rm{CF}$ with $r_\parallel$
parallel to the direction of interval-averaged (non-local) magnetic field
\citep{Matthaeus1990JGR, Dasso2005ApJ}, is not so prominent in our cases.

\begin{figure}
\centering
\includegraphics[width=17cm]{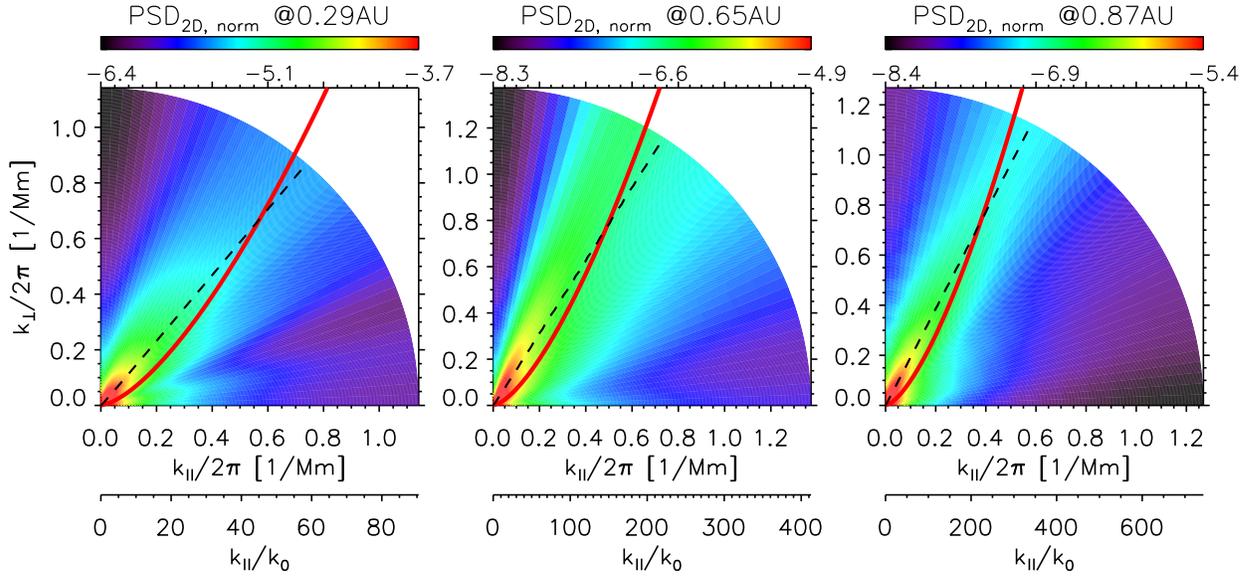}
\caption{$\rm{PSD}_{\rm{2D}}(k_\parallel,k_\bot)$ at three positions (0.29, 0.65, and 0.87~AU)
as derived from $\rm{CF}_{\rm{2D}}(r_\parallel,r_\bot)$ according to Equation~\ref{Eq.7}
following the projection-slice theorem. The major components (ridge distribution
with its centroid position aligned as black dashed line) may be roughly described by Equation~\ref{Eq.10} shown as red solid line,
which means the wave-vector anisotropy develops as the outer-scale wave-number ($k_0$)
becomes smaller with increasing heliocentric distance. A weakening trend of the minor
component that is inclined to $k_\parallel$ becomes also visible.}
\label{Fig.5}
\end{figure}

Ideally, the corresponding $\rm{PSD}(k_\parallel, k_\bot)$ can be gained
directly from 2D Fourier transform of $\rm{CF}_{\rm{2D}}(r_\parallel,
r_\bot)$. However, in practice, the transformed value might be negative or
not certainly positive, thereby restraining the application of the direct 2D
Fourier transform. To obtain $\rm{PSD}_{\rm{2D}}(k_\parallel, k_\bot)$, we
then turn to Equation~\ref{Eq.7} for a step-by-step derivation. Firstly, by
integrating $\rm{CF}_{\rm{2D}}(r_\parallel, r_\bot)$ over the path normal to
the direction with certain angle $\theta'$ with respect to $r_\parallel$, the
reduced 1D $\rm{CF}_{\rm{1D}}(r)$ corresponding to the angle $\theta'$ is
calculated. Secondly, the corresponding $\rm{PSD}$ as a Fourier transform of
$\rm{CF}_{\rm{1D}}$ is calculated. To guarantee the positivity of the
estimated $\rm{PSD}$, $\rm{CF}_{\rm{1D}}$ is fitted before transformation
with a function related to that for $\rm{SF}$ as previously described.
According to the projection-slice theorem, the estimated $\rm{PSD}$ profile
is essentially a slice of $\rm{PSD}_{\rm{2D}}(k_\parallel, k_\bot)$ along
$\mathbf{k}$ with $\theta'$ with respect to $k_\parallel$. Thirdly,
$\rm{PSD}_{\rm{2D}}(k_\parallel,k_\bot)$ is formed by assembling various
$\rm{PSD}$ profiles, with different angles ranging from $0^\circ$ to
$90^\circ$ with respect to $k_\parallel$. We note that, in calculation,
$\rm{CF}_{\rm{2D}}$ and $\rm{CF}_{\rm{1D}}$ one cannot let $r$ go to
infinity. As a result, the transformed $\rm{PSD}_{\rm{2D}}$ may slightly
depart from the real one. Therefore, in Figure~\ref{Fig.5}, we just present
the normalized $\rm{PSD}_{\rm{2D,n}}$ rather than the absolute
$\rm{PSD}_{\rm{2D}}$. The uncertainty (confidence interval) for the estimated
$\rm{PSD}_{\rm{2D}}$ is not provided here, since due to the complexity of the
estimation method that was not yet possible.

Obviously, the normalized $\rm{PSD}_{\rm{2D,n}}$ shown in Figure~\ref{Fig.5}
is not uniformly distributed at all angles, indicating an anisotropic
wave-vector distribution. This anisotropy is mainly characterized by a ridge
distribution which has a bias towards $k_\bot$ as compared to $k_\parallel$.
Moreover, as the heliographic distance increases, the ridge distribution
becomes more inclined toward $k_\bot$ at the same $|k|$, in association with
lower PSD (darker blue in the figure) around the $k_\parallel$ region and
higher PSD (brighter blue in the figure) around the $k_\bot$ region. The
discovery of this bent ridge and its radial evolution imply that solar wind
turbulent energy cascades preferentially along the $k_\bot$ as compared to
the $k_\parallel$ axis, and the turbulence cascade radially develops with
more energy cascading to the $k_\bot$ region, as the scale ($1/|k|$) is
shifting away from the radially-growing outer scale ($1/|k_0|$). In addition
to the major ridge distribution, a minor population seems to exist close to
$k_\parallel$ (see Figure~\ref{Fig.5}a), and appears to become weaker at
farther distances (see Figures~\ref{Fig.5}b,c). The observational fact that
$\rm{PSD}$ is composed of two populations, with the major one bending more
perpendicularly and the minor one becoming weaker, seems compatible with the
previously suggested two-component turbulence model, which invokes
non-damping convective structures (spatially varying across $\mathbf{B}_0$)
that are superposed on damping Alfv\'en waves (spatially varying along
$\mathbf{B}_0$) \citep{Tu1993JGR}.

To emphasize the trend of the ridge distribution and its radial evolution, we estimate the ridge position of every scale by averaging the angles with local lg(PSD) as the weights (i.e., first-order moment centroid method). The estimated ridge positions are shown as black dashed lines in Figure~\ref{Fig.5} and appear straight. Whether or not the straightness is realistic is yet unknown. Furthermore, we fit the estimated ridge position with following
simple formula,
\begin{equation}
k_ \parallel   = \alpha \cdot k_0^{1/3}  \cdot k_\bot ^{2/3},\label{Eq.10}
\end{equation}
where $k_0(=2\pi/(V_{\rm{sw}}\cdot\tau_c))$ is related to the outer-scale
wave-number. $\alpha$ is the coefficient to be fitted, which is $\sim$ 3.2, 3.9, and 3.9 for our three
cases. We find that the simple $k_\parallel\rm{-}k_\bot$-relation profile
according to Equation~\ref{Eq.10} is basically coincident with the observed
ridge distribution. However, some departures, e.g., the estimated black dashed line looks
more straight than the fitted red line, still remain. Nevertheless, the
relation ($k_\parallel \sim k_0^{1/3}k_\bot ^{2/3}$), as predicted by the
critical-balance hypothesis \citep{Goldreich1995ApJ} for MHD turbulence,
seems to describe well the observed anisotropy of solar wind turbulence. The
role of $k_0$, which was once neglected in previous observational studies, in
shaping the anisotropy shall be emphasized here. It may be the reduction in
$k_0$ which causes the development of the spectral anisotropy (increasing
inclination toward $k_\bot$ at the same $|k|$) in interplanetary space as
heliographic distance increases.

\begin{figure}
\centering
\includegraphics[width=17cm]{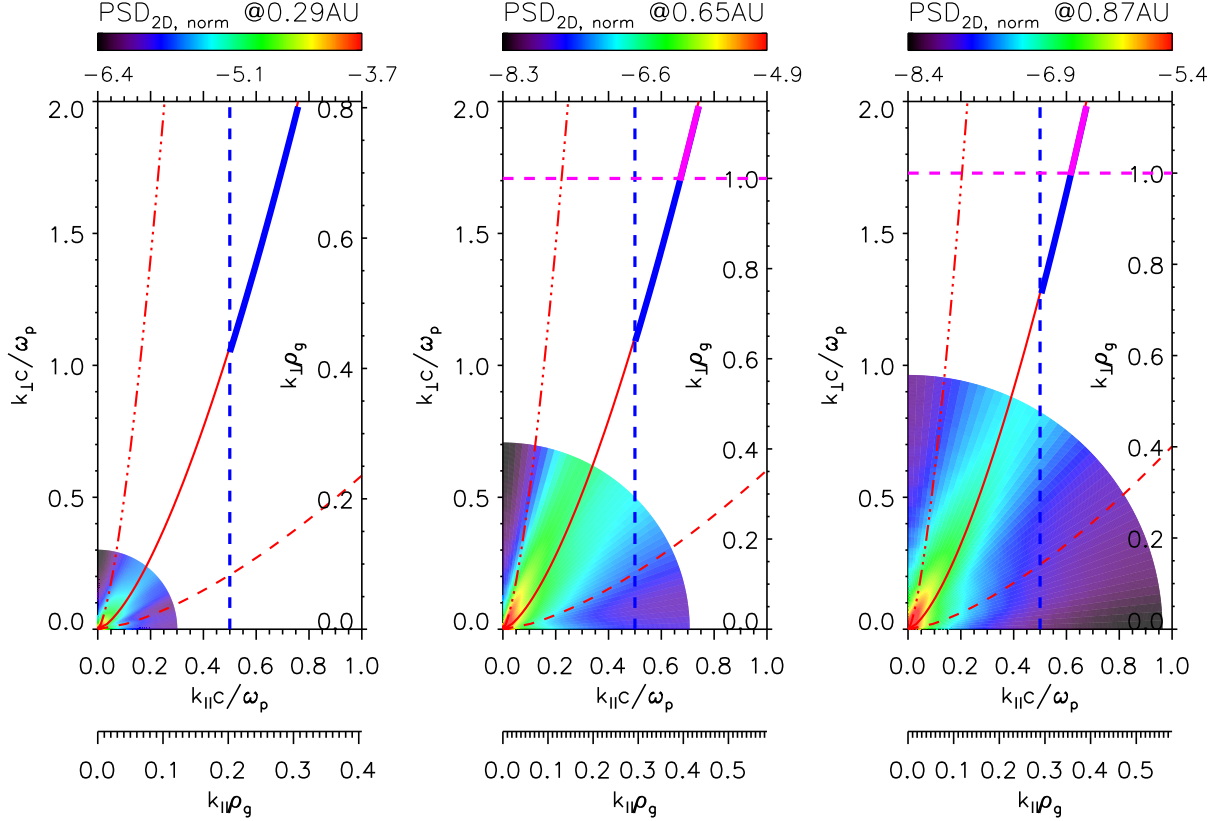}
\caption{Implication of solar wind heating mechanism from the extension of ridge profile. Red solid lines denote the extended ridge profiles ($k_\parallel=\alpha\cdot k_0^{1/3} \cdot k_\bot^{2/3}$ with $\alpha=3.2, 3.9, 3.9$ for the three panels) as superposed on $\rm{PSD}_{\rm{2D}}(k_\parallel,k_\bot)$ in larger wave-vector space. Large $k_\parallel$ with $k_\parallel c/\omega_{\rm{p}}>0.5$ (to the right of blue dashed line) indicates the region of ion cyclotron resonance. Landau resonance becomes active when $k_\bot\rho_{\rm{g}}>1$ (above magenta dashed line) and dominant for $k_\parallel c/\omega_{\rm{p}}\ll1.0$. Radial evolution of the approximated ridge profile and its intersections with the threshold lines ($k_\parallel c/\omega_{\rm{p}}=0.5$ and $k_\bot\rho_{\rm{g}}=1$) indicate the transition of cascade termination from cyclotron resonance to Landau resonance as the solar wind flows further away. Ridge profiles with larger (smaller) $\alpha(=10(1.1))$ (red dashed (red dash-dot-dot) lines) would lead to cyclotron resonance (Landau resonance) separately at all distances without a transition, which seems unrealistic. The proton gyroradius (proton inertial length)
$\rho_{\rm{g}} (c/\omega_{\rm{p}})$ is 17(42), 48(82), and 70(121)~km at 0.29, 0.65, and 0.87~AU.}
\label{Fig.6}
\end{figure}

Solar wind heating mechanism may be inferred from the radial evolution of the ridge trend. According to linear Vlasov theory, Alfv\'en waves with plasma $\beta_{\rm{p}}\in[0.1,1.0]$ usually become dissipated due to proton cyclotron resonance when they have $k_\parallel c/\omega_{\rm{p}}\in[0.1,1.0]$, where $c/\omega_{\rm{p}}$ is the proton inertial length  \citep{Gary2004JGRA}. On the other hand, Landau resonance becomes more and more prominent as plasma $\beta_p$ rises \citep{Gary2004JGRA} and $k_\bot\rho_{\rm{g}}$ increases \citep{Howes2006ApJ}, where $\rho_{\rm{g}}$ is the proton gyroradius. \citet{Howes2011ApJ} pointed out that, Landau damping calculated in the gyro-kinetic limit is not sufficient for the empirically estimated proton heating \citep{Cranmer2009ApJ} at small heliocentric distances ($R<0.8$~AU). At these small distances, turbulent cascade is speculated to approach to proton cyclotron frequency before being terminated by Landau resonance \citep{Howes2011ApJ}. However, the relative contributions from cyclotron resonance and Landau resonance to solar wind heating at different radial distances have not yet been addressed from observations. The approximated relation ($k_ \parallel=\alpha\cdot k_0^{1/3} \cdot k_\bot^{2/3}$ with $\alpha\in[3, 4]$) for the observed ridge distribution may be used to address this issue. In Figure~\ref{Fig.6}, we just simply extend the approximated ridge profile in a larger wave-vector space to see what kind of resonance (cyclotron or Landau) would probably terminate the cascade. As a result, we find that, at the three distances within 1~AU, the extended ridge profile clearly approaches to cyclotron resonance (marked by $k_\parallel c/\omega_{\rm{p}}>0.5$) ahead of Landau resonance (marked by $k_\bot\rho_{\rm{g}}>1.0$). Moreover, one may expect that, as the distance increases ($\geq1$~AU) in association with reduction of $k_0$, the ridge profile would first exceed the threshold $k_\bot\rho_{\rm{g}}=1$ before approaching to $k_\parallel c/\omega_p=0.5$, which implies a dominance of Landau damping over cyclotron damping at larger distances. According to critical-balance theory in MHD and kinetic regimes\citep{Schekochihin2009ApJS}, the extension of ridge in the MHD inertial range may be still related to Equation~\ref{Eq.10}, while the extension part in the kinetic (dissipation) range may deviate from Equation~\ref{Eq.10} with more inclination towards $k_\bot$. For sake of simplicity, we neglect such deviation of extended ridge in the kinetic range from that in the inertial range.

The value of coefficient $\alpha$ is also worth emphasizing here. If $\alpha$ were one third of the approximated value ($3.3/3=1.1$), the ridge profile (red dash-dot-dot line in Figure~\ref{Fig.6}) would exceed $k_\bot\rho_{\rm{g}}=1$ without approaching to $k_\parallel c/\omega_{\rm{p}}=0.5$, leading to insufficient heating rate by Landau resonance within 1~AU according to the gyro-kinetic prescription by \citet{Howes2011ApJ}. On the other side, if $\alpha$ were too large (saying $3.3\times3=10$), the ridge profile (red dashed line in Figure~\ref{Fig.6}) would lie well below $k_\bot\rho_g=1$, implying the absence of transition from cyclotron damping to Landau damping around 1~AU (inconsistent with the conclusion by \citet{Howes2011ApJ}). Therefore, the $\alpha$ value ($\in[3,4]$ obtained here) besides $k_0$ is another important parameter for grasping the essence of solar wind heating mechanism. $\alpha$ may be expressed as the ratio of $\varepsilon$ to $V_{\rm{A}}^3 k_{\rm{0}}$ with $\varepsilon$ being the energy cascade rate if $k_\parallel=(\varepsilon/V_{\rm{A}}^3)^{1/3}k_\bot^{2/3}$, which is usually assumed in critical-balance theory \citep{Goldreich1995ApJ, Schekochihin2009ApJS}.

\section{Summary and discussion}

We have made the first successful attempt to reconstruct, on the basis of
single spacecraft measurements, the 2D spectral density
$\rm{PSD}_{\rm{2D}}(k_\parallel,k_\bot)$ for solar wind MHD turbulence. We
estimate the angular distribution of the second-order structure function
$\rm{SF}(\tau, \theta_{\rm{RB}})$, and derive the corresponding correlation
function $\rm{CF}_{\rm{2D}}(r_\parallel, r_\bot)$, which in principle is an
inverse 2D Fourier transform of $\rm{PSD}_{\rm{2D}}(k_\parallel, k_\bot)$.
The transformation from time scale $\tau$ to spatial scale $r$, when building
up $\rm{CF}_{\rm{2D}}(r_\parallel, r_\bot)$, is based on Taylor's hypothesis
that solar wind fluctuations are quasi-stationary within the flow transit
time scale, as the solar wind passes by the spacecraft. The 2D direct Fourier
transform of $\rm{CF}_{\rm{2D}}(r_\parallel,r_\bot)$ fails to guarantee the
required positivity of $\rm{PSD}_{\rm{2D}}(k_\parallel,k_\bot)$.
Alternatively, we employ for the first time a method based on the
projection-slice theorem, which connects the integrated
$\rm{CF}_{\rm{1D}}(r,\theta')$ with the corresponding slice
$\rm{PSD}_{\rm{2D}}(k,\theta')$ of the $\rm{PSD}$ via a 1D Fourier transform,
to fulfill that task. Before the 1D Fourier transformation,
$\rm{CF}_{\rm{1D}}(r,\theta')$ is fitted smoothly to guarantee the positivity
of the transformed $\rm{PSD}_{\rm{2D}}(k,\theta')$.

As a result, $\rm{SF}(\tau, \theta_{\rm{RB}})$ shows a non-uniform angular
distribution with more power being located in the perpendicular region
($\theta_{\rm{RB}}\sim90^\circ$) than in the parallel region
($\theta_{\rm{RB}}\sim0^\circ$) of wave-vector space. Moreover, there is
angular dependence of the scaling law for $\rm{SF}(\tau, \theta_{\rm{RB}})$
at short scales, whereby the scaling index $p$ drops from $\sim0.9$ at
$\theta_{\rm{RB}}=0^\circ$ to $\sim0.6$ at $\theta_{\rm{RB}}=90^\circ$. We
find that $\rm{SF}(\tau,\theta_{\rm{RB}})$ have at all three positions (0.29,
0.65, and 0.87~AU) the two above properties, indicating the prevalence of
anisotropy in the turbulence throughout the inner heliosphere. This result
obtained within 1~AU is similar to that found for the $\rm{SF}$ anisotropy
beyond 1~AU \citep{Luo2010ApJ}. The corresponding correlation functions
$\rm{CF}_{\rm{2D}}(r_\parallel,r_\bot)$ clearly show that magnetic
fluctuations are correlated at longer (shorter) length along (across) the
background magnetic field.

The corresponding $\rm{PSD}_{\rm{2D}}(k_\parallel,k_\bot)$ at the positions
within 1~AU is revealed to have a ridge distribution with a bias towards
$k_\bot$ as compared to $k_\parallel$, suggesting a preferential cascading
along $k_\bot$. This kind of ridge distribution has never been reported in
previous studies at 1~AU, e.g. those based on the wave-telescope
(k-filtering) method \citep{Narita2010PhRvL, Sahraoui2010PhRvL}. Furthermore,
this ridge distribution is found to become ever more inclined toward the
$k_\bot$ axis with increasing heliographic distance, thus indicating a radial
development of the wave-vector anisotropy. The observed radial evolution of
the ridge casts new light on the scaling relation between $k_\parallel$ and
$k_\bot$, which may empirically be approximated by $k_\parallel\simeq \alpha
k_0^{1/3} k_\bot ^{2/3}$, with $\alpha\in[3, 4]$ and $k_0$ being the
wave-number of the outer scale. This approximation for the wave-vector
anisotropy seems to indicate critical-balance-type cascading
\citep{Goldreich1995ApJ} of solar wind turbulence. A possible influence of
$k_0$ on the anisotropy development, which was neglected in previous
observational analyses, is also found.

However, the evolution of the ridge distribution cannot represent the whole
story about wave-vector anisotropy of solar wind turbulence. There seems to
be a minor population located near $k_\parallel$, which is beyond the scope
of critical-balance turbulence theory. The apparent two-component
distribution of $\rm{PSD}_{\rm{2D}}(k_\parallel,k_\bot)$ seems to be
connected with previous two-component models, e.g., models with
``slab''+``2D'' \citep{Matthaeus1990JGR}, models composed of Alfv\'en
waves and convected structures \citep{Tu1993JGR}, and conjectures with critical-balanced component plus slab component \citep{Forman2011ApJ, He2012bApJ}. We find that the minor
population seems to weaken further with increasing heliographic distance,
leaving more energy distributed in the region close to the $k_\bot$ axis.
This gradual migration of energy towards $k_\bot$ might indicate a relative
enrichment of turbulence energy carried by convective structures and explain
the observed associated shortage of Alfv\'enicity, which was already
discussed in the previous two-component model by \cite{Tu1993JGR}. The observed ``slab''-like minor component is crucial for scattering of energetic particles in the interplanetary space \citep{Bieber1996JGR, Chandran2000ApJ, Qin2002ApJ}. The radial evolution of anisotropic turbulence may be quantified in the future and incorporated into the transport model of energetic particles.

The estimated $\rm{PSD}_{\rm{2D}}(k_\parallel,k_\bot)$ is believed to impose
valuable observational constraints on the theoretical models of solar wind
turbulence. Recently, \citet{Cranmer2012ApJ} modelled
$\rm{PSD}_{\rm{2D}}(k_\parallel, k_\bot)$ at different heliocentric distances
by solving a set of 2D cascade-advection-diffusion equations, with the total
power pre-determined by the damped wave-action conservation equation and the
reduced $\rm{PSD}_{\rm{1D}}(k_\bot)$ pre-set by the 1D advection-diffusion
equation. Their modelled $\rm{PSD}_{\rm{2D}}(k_\parallel,k_\bot)$ looks
partly similar to our observational spectrum, in the sense of where the major
power is located. However, the differences in distribution pattern and radial
evolution between observational and modelled spectra call for a substantial
improvement of the models for solar wind turbulence.

The approximated ridge profile as extended to large $k_\parallel$ and $k_\bot$ may give a hint about the resonance type responsible for solar wind heating at different radial distances. The extended ridge profile at small distances ($R<0.8$~AU) is found to reach larger $k_\parallel$ where proton cyclotron resonance acts before Landau damping sets in. As the distance increases, the extended ridge profile, which is inclined more towards $k_\bot$ due to the reduction of $k_0$, tends to arrive at Landau resonance before cyclotron resonance. Such performance of the approximated ridge profile confirms observationally previous conjecture about the transition from cyclotron resonance to Landau resonance with increasing heliographic distance \citep{Howes2011ApJ}.

Our results are just limited to the MHD inertial range, but the analysis
should be extended to kinetic scales where several typical properties have
been revealed: steeper power-law magnetic spectrum \citep{Sahraoui2009PhRvL,
Alexandrova2009PhRvL}, enhanced electric-field spectrum
\citep{Bale2005PhRvL}, enhanced magnetic compressibility \citep{Smith2006ApJ,
Hamilton2008JGRA, Salem2012ApJ, He2012aApJ}, and two-component pattern in the
magnetic helicity \citep{He2011ApJ, Podesta2011ApJ, He2012aApJ, He2012bApJ}.
These observations seem to be in favour of the oblique Alfv\'en waves or
kinetic Alfv\'en waves (KAW) as the candidate for explaining the dominant
fluctuations in ion-scale turbulence. The oblique Alfv\'en/ion-cyclotron
waves may be via resonance diffusion \citep{Marsch2011AnGeo} responsible for
the formation of the observed wide proton beam. The theory of KAW itself and
its role in kinetic turbulence have been studied intensively
\citep{Hollweg1999JGR, Wu2004NPGeo, Howes2008JGRA, Zhao2011PhPl,
Voitenko2011NPGeo, Howes2011PhRvL}. Ion cyclotron waves, which are considered
responsible for the ion perpendicular heating \citep{Bourouaine2010GeoRL},
were also identified \citep{Jian2009ApJ, He2011ApJ}. There are other possible
wave modes, e.g. fast whistler waves, ion Bernstein waves, and fast-cyclotron
waves, which may exist in kinetic turbulence \citep{Gary2012ApJ,
TenBarge2012ApJ, Xiong2012SoPh}. Spectral break at the ion-kinetic scale seems to be almost constant (about 0.5~Hz in the spacecraft frame) with radial distance \citep{Perri2010ApJ, Bourouaine2012ApJ}. The spectral break frequency might corresponds to the proton inertial length in quasi-2D turbulence when considering a large-scale background magnetic field $\mathbf{B}_0$ (which is obtained through averaging over a time period higher than 1 hour) \citep{Bourouaine2012ApJ}. However, not much is presently known about
the radial evolution of solar wind turbulence at ion-kinetic scales.

In the future, with the help of high-time-resolution measurements to be made
by the wave and particle instruments flown on such mission like Solar Orbiter
and Solar Probe Plus, the radial evolution of the wave-vector anisotropy at
kinetic scale may be studied, and more new results will be obtained on the
spectrum anisotropy in the inertial range that was analysed here.

\begin{acknowledgements}
{\bf Acknowledgements:} This work was supported by the National Natural
Science Foundation of China under Contract Nos. 41174148, 41222032, 40890162,
40931055, and 41231069. JS He appreciates helpful discussions with J.-S.
Zhao, R. Wicks, and Y. Voitenko.
\end{acknowledgements}

\bibliographystyle{apj}
\bibliography{references}

\begin{thebibliography}{76}
\expandafter\ifx\csname natexlab\endcsname\relax\def\natexlab#1{#1}\fi

\bibitem[{{Alexandrova} {et~al.}(2009){Alexandrova}, {Saur}, {Lacombe},
  {Mangeney}, {Mitchell}, {Schwartz}, \& {Robert}}]{Alexandrova2009PhRvL}
{Alexandrova}, O., {Saur}, J., {Lacombe}, C., {Mangeney}, A., {Mitchell}, J.,
  {Schwartz}, S.~J., \& {Robert}, P. 2009, Physical Review Letters, 103, 165003

\bibitem[{{Bale} {et~al.}(2005){Bale}, {Kellogg}, {Mozer}, {Horbury}, \&
  {Reme}}]{Bale2005PhRvL}
{Bale}, S.~D., {Kellogg}, P.~J., {Mozer}, F.~S., {Horbury}, T.~S., \& {Reme},
  H. 2005, Physical Review Letters, 94, 215002

\bibitem[{{Bavassano} {et~al.}(1982){Bavassano}, {Dobrowolny}, {Mariani}, \&
  {Ness}}]{Bavassano1982JGR}
{Bavassano}, B., {Dobrowolny}, M., {Mariani}, F., \& {Ness}, N.~F. 1982, \jgr,
  87, 3617

\bibitem[{{Beresnyak} \& {Lazarian}(2008)}]{Beresnyak2008ApJ}
{Beresnyak}, A., \& {Lazarian}, A. 2008, \apj, 682, 1070

\bibitem[{{Bieber} {et~al.}(1996){Bieber}, {Wanner}, \&
  {Matthaeus}}]{Bieber1996JGR}
{Bieber}, J.~W., {Wanner}, W., \& {Matthaeus}, W.~H. 1996, \jgr, 101, 2511

\bibitem[{{Biskamp} \& {M{\"u}ller}(2000)}]{Biskamp2000PhPl}
{Biskamp}, D., \& {M{\"u}ller}, W.-C. 2000, Physics of Plasmas, 7, 4889

\bibitem[{{Boldyrev}(2005)}]{Boldyrev2005ApJ}
{Boldyrev}, S. 2005, \apjl, 626, L37

\bibitem[{{Bourouaine} {et~al.}(2012){Bourouaine}, {Alexandrova}, {Marsch}, \&
  {Maksimovic}}]{Bourouaine2012ApJ}
{Bourouaine}, S., {Alexandrova}, O., {Marsch}, E., \& {Maksimovic}, M. 2012,
  \apj, 749, 102

\bibitem[{{Bourouaine} {et~al.}(2010){Bourouaine}, {Marsch}, \&
  {Neubauer}}]{Bourouaine2010GeoRL}
{Bourouaine}, S., {Marsch}, E., \& {Neubauer}, F.~M. 2010, \grl, 37, 14104

\bibitem[{{Bovik}(2000)}]{Bovik2000}
{Bovik}, A.~C. 2000, {Handbook of Image and Video Processing}, ed. {Bovik,
  A.~C.}

\bibitem[{{Bruno} \& {Carbone}(2005)}]{Bruno2005LRSP}
{Bruno}, R., \& {Carbone}, V. 2005, Living Reviews in Solar Physics, 2, 4

\bibitem[{{Chandran}(2000)}]{Chandran2000ApJ}
{Chandran}, B.~D.~G. 2000, \apj, 529, 513

\bibitem[{{Chandran}(2008)}]{Chandran2008ApJ}
---. 2008, \apj, 685, 646

\bibitem[{{Chen} {et~al.}(2010){Chen}, {Horbury}, {Schekochihin}, {Wicks},
  {Alexandrova}, \& {Mitchell}}]{Chen2010PhRvL}
{Chen}, C.~H.~K., {Horbury}, T.~S., {Schekochihin}, A.~A., {Wicks}, R.~T.,
  {Alexandrova}, O., \& {Mitchell}, J. 2010, Physical Review Letters, 104,
  255002

\bibitem[{{Chen} {et~al.}(2012){Chen}, {Mallet}, {Schekochihin}, {Horbury},
  {Wicks}, \& {Bale}}]{Chen2012ApJ}
{Chen}, C.~H.~K., {Mallet}, A., {Schekochihin}, A.~A., {Horbury}, T.~S.,
  {Wicks}, R.~T., \& {Bale}, S.~D. 2012, \apj, 758, 120

\bibitem[{{Cho} {et~al.}(2002){Cho}, {Lazarian}, \& {Vishniac}}]{Cho2002ApJ}
{Cho}, J., {Lazarian}, A., \& {Vishniac}, E.~T. 2002, \apj, 564, 291

\bibitem[{{Cranmer} {et~al.}(2009){Cranmer}, {Matthaeus}, {Breech}, \&
  {Kasper}}]{Cranmer2009ApJ}
{Cranmer}, S.~R., {Matthaeus}, W.~H., {Breech}, B.~A., \& {Kasper}, J.~C. 2009,
  \apj, 702, 1604

\bibitem[{{Cranmer} \& {van Ballegooijen}(2012)}]{Cranmer2012ApJ}
{Cranmer}, S.~R., \& {van Ballegooijen}, A.~A. 2012, \apj, 754, 92

\bibitem[{{Dasso} {et~al.}(2005){Dasso}, {Milano}, {Matthaeus}, \&
  {Smith}}]{Dasso2005ApJ}
{Dasso}, S., {Milano}, L.~J., {Matthaeus}, W.~H., \& {Smith}, C.~W. 2005,
  \apjl, 635, L181

\bibitem[{{Forman} {et~al.}(2011){Forman}, {Wicks}, \&
  {Horbury}}]{Forman2011ApJ}
{Forman}, M.~A., {Wicks}, R.~T., \& {Horbury}, T.~S. 2011, \apj, 733, 76

\bibitem[{{Gary} {et~al.}(2012){Gary}, {Chang}, \& {Wang}}]{Gary2012ApJ}
{Gary}, S.~P., {Chang}, O., \& {Wang}, J. 2012, \apj, 755, 142

\bibitem[{{Gary} \& {Nishimura}(2004)}]{Gary2004JGRA}
{Gary}, S.~P., \& {Nishimura}, K. 2004, Journal of Geophysical Research (Space
  Physics), 109, 2109

\bibitem[{{Goldreich} \& {Sridhar}(1995)}]{Goldreich1995ApJ}
{Goldreich}, P., \& {Sridhar}, S. 1995, \apj, 438, 763

\bibitem[{{Goldstein} {et~al.}(1995){Goldstein}, {Roberts}, \&
  {Matthaeus}}]{Goldstein1995ARA&A}
{Goldstein}, M.~L., {Roberts}, D.~A., \& {Matthaeus}, W.~H. 1995, Ann. Rev.
  Astron. \& Astrophys., 33, 283

\bibitem[{{Grappin} {et~al.}(1990){Grappin}, {Mangeney}, \&
  {Marsch}}]{Grappin1990JGR}
{Grappin}, R., {Mangeney}, A., \& {Marsch}, E. 1990, \jgr, 95, 8197

\bibitem[{{Hamilton} {et~al.}(2008){Hamilton}, {Smith}, {Vasquez}, \&
  {Leamon}}]{Hamilton2008JGRA}
{Hamilton}, K., {Smith}, C.~W., {Vasquez}, B.~J., \& {Leamon}, R.~J. 2008,
  Journal of Geophysical Research (Space Physics), 113, 1106

\bibitem[{{He} {et~al.}(2011){He}, {Marsch}, {Tu}, {Yao}, \&
  {Tian}}]{He2011ApJ}
{He}, J., {Marsch}, E., {Tu}, C., {Yao}, S., \& {Tian}, H. 2011, \apj, 731, 85

\bibitem[{{He} {et~al.}(2012{\natexlab{a}}){He}, {Tu}, {Marsch}, \&
  {Yao}}]{He2012aApJ}
{He}, J., {Tu}, C., {Marsch}, E., \& {Yao}, S. 2012{\natexlab{a}}, \apjl, 745,
  L8

\bibitem[{{He} {et~al.}(2012{\natexlab{b}}){He}, {Tu}, {Marsch}, \&
  {Yao}}]{He2012bApJ}
---. 2012{\natexlab{b}}, \apj, 749, 86

\bibitem[{{Hollweg}(1974)}]{Hollweg1974JGR}
{Hollweg}, J.~V. 1974, \jgr, 79, 1539

\bibitem[{{Hollweg}(1999)}]{Hollweg1999JGR}
---. 1999, \jgr, 104, 14811

\bibitem[{{Horbury} {et~al.}(2008){Horbury}, {Forman}, \&
  {Oughton}}]{Horbury2008PhRvL}
{Horbury}, T.~S., {Forman}, M., \& {Oughton}, S. 2008, Physical Review Letters,
  101, 175005

\bibitem[{{Howes}(2011)}]{Howes2011ApJ}
{Howes}, G.~G. 2011, \apj, 738, 40

\bibitem[{{Howes} {et~al.}(2006){Howes}, {Cowley}, {Dorland}, {Hammett},
  {Quataert}, \& {Schekochihin}}]{Howes2006ApJ}
{Howes}, G.~G., {Cowley}, S.~C., {Dorland}, W., {Hammett}, G.~W., {Quataert},
  E., \& {Schekochihin}, A.~A. 2006, \apj, 651, 590

\bibitem[{{Howes} {et~al.}(2008){Howes}, {Cowley}, {Dorland}, {Hammett},
  {Quataert}, \& {Schekochihin}}]{Howes2008JGRA}
---. 2008, Journal of Geophysical Research (Space Physics), 113, 5103

\bibitem[{{Howes} {et~al.}(2011){Howes}, {Tenbarge}, {Dorland}, {Quataert},
  {Schekochihin}, {Numata}, \& {Tatsuno}}]{Howes2011PhRvL}
{Howes}, G.~G., {Tenbarge}, J.~M., {Dorland}, W., {Quataert}, E.,
  {Schekochihin}, A.~A., {Numata}, R., \& {Tatsuno}, T. 2011, Physical Review
  Letters, 107, 035004

\bibitem[{{Jian} {et~al.}(2009){Jian}, {Russell}, {Luhmann}, {Strangeway},
  {Leisner}, \& {Galvin}}]{Jian2009ApJ}
{Jian}, L.~K., {Russell}, C.~T., {Luhmann}, J.~G., {Strangeway}, R.~J.,
  {Leisner}, J.~S., \& {Galvin}, A.~B. 2009, \apjl, 701, L105

\bibitem[{{Lithwick} {et~al.}(2007){Lithwick}, {Goldreich}, \&
  {Sridhar}}]{Lithwick2007ApJ}
{Lithwick}, Y., {Goldreich}, P., \& {Sridhar}, S. 2007, \apj, 655, 269

\bibitem[{{Luo} \& {Wu}(2010)}]{Luo2010ApJ}
{Luo}, Q.~Y., \& {Wu}, D.~J. 2010, \apjl, 714, L138

\bibitem[{{Marsch}(2006)}]{Marsch2006LRSP}
{Marsch}, E. 2006, Living Reviews in Solar Physics, 3, 1

\bibitem[{{Marsch} \& {Bourouaine}(2011)}]{Marsch2011AnGeo}
{Marsch}, E., \& {Bourouaine}, S. 2011, Annales Geophysicae, 29, 2089

\bibitem[{{Marsch} \& {Tu}(1989)}]{Marsch1989JPlPh}
{Marsch}, E., \& {Tu}, C.-Y. 1989, Journal of Plasma Physics, 41, 479

\bibitem[{{Marsch} \& {Tu}(1990)}]{Marsch1990JGR}
---. 1990, \jgr, 95, 8211

\bibitem[{{Matthaeus} {et~al.}(1990){Matthaeus}, {Goldstein}, \&
  {Roberts}}]{Matthaeus1990JGR}
{Matthaeus}, W.~H., {Goldstein}, M.~L., \& {Roberts}, D.~A. 1990, \jgr, 95,
  20673

\bibitem[{{M{\"u}ller} {et~al.}(2003){M{\"u}ller}, {Biskamp}, \&
  {Grappin}}]{Mueller2003PhRvE}
{M{\"u}ller}, W.-C., {Biskamp}, D., \& {Grappin}, R. 2003, \pre, 67, 066302

\bibitem[{{Narita} {et~al.}(2010){Narita}, {Glassmeier}, {Sahraoui}, \&
  {Goldstein}}]{Narita2010PhRvL}
{Narita}, Y., {Glassmeier}, K.-H., {Sahraoui}, F., \& {Goldstein}, M.~L. 2010,
  Physical Review Letters, 104, 171101

\bibitem[{{N{\'e}meth} {et~al.}(2010){N{\'e}meth}, {Facsk{\'o}}, \&
  {Lucek}}]{Nemeth2010SoPh}
{N{\'e}meth}, Z., {Facsk{\'o}}, G., \& {Lucek}, E.~A. 2010, \solphys, 266, 149

\bibitem[{{Osman} \& {Horbury}(2007)}]{Osman2007ApJ}
{Osman}, K.~T., \& {Horbury}, T.~S. 2007, \apjl, 654, L103

\bibitem[{{Perri} {et~al.}(2010){Perri}, {Carbone}, \& {Veltri}}]{Perri2010ApJ}
{Perri}, S., {Carbone}, V., \& {Veltri}, P. 2010, \apjl, 725, L52

\bibitem[{{Pincon} \& {Lefeuvre}(1991)}]{Pincon1991JGR}
{Pincon}, J.~L., \& {Lefeuvre}, F. 1991, \jgr, 96, 1789

\bibitem[{{Podesta}(2009)}]{Podesta2009ApJ}
{Podesta}, J.~J. 2009, \apj, 698, 986

\bibitem[{{Podesta} \& {Bhattacharjee}(2010)}]{Podesta2010ApJ}
{Podesta}, J.~J., \& {Bhattacharjee}, A. 2010, \apj, 718, 1151

\bibitem[{{Podesta} \& {Gary}(2011)}]{Podesta2011ApJ}
{Podesta}, J.~J., \& {Gary}, S.~P. 2011, \apj, 734, 15

\bibitem[{{Qin} {et~al.}(2002){Qin}, {Matthaeus}, \& {Bieber}}]{Qin2002ApJ}
{Qin}, G., {Matthaeus}, W.~H., \& {Bieber}, J.~W. 2002, \apjl, 578, L117

\bibitem[{{Roberts} {et~al.}(1987){Roberts}, {Goldstein}, {Klein}, \&
  {Matthaeus}}]{Roberts1987JGR}
{Roberts}, D.~A., {Goldstein}, M.~L., {Klein}, L.~W., \& {Matthaeus}, W.~H.
  1987, \jgr, 92, 12023

\bibitem[{{Sahraoui} {et~al.}(2010){Sahraoui}, {Goldstein}, {Belmont}, {Canu},
  \& {Rezeau}}]{Sahraoui2010PhRvL}
{Sahraoui}, F., {Goldstein}, M.~L., {Belmont}, G., {Canu}, P., \& {Rezeau}, L.
  2010, Physical Review Letters, 105, 131101

\bibitem[{{Sahraoui} {et~al.}(2009){Sahraoui}, {Goldstein}, {Robert}, \&
  {Khotyaintsev}}]{Sahraoui2009PhRvL}
{Sahraoui}, F., {Goldstein}, M.~L., {Robert}, P., \& {Khotyaintsev}, Y.~V.
  2009, Physical Review Letters, 102, 231102

\bibitem[{{Salem} {et~al.}(2012){Salem}, {Howes}, {Sundkvist}, {Bale},
  {Chaston}, {Chen}, \& {Mozer}}]{Salem2012ApJ}
{Salem}, C.~S., {Howes}, G.~G., {Sundkvist}, D., {Bale}, S.~D., {Chaston},
  C.~C., {Chen}, C.~H.~K., \& {Mozer}, F.~S. 2012, \apjl, 745, L9

\bibitem[{{Schekochihin} {et~al.}(2009){Schekochihin}, {Cowley}, {Dorland},
  {Hammett}, {Howes}, {Quataert}, \& {Tatsuno}}]{Schekochihin2009ApJS}
{Schekochihin}, A.~A., {Cowley}, S.~C., {Dorland}, W., {Hammett}, G.~W.,
  {Howes}, G.~G., {Quataert}, E., \& {Tatsuno}, T. 2009, \apjs, 182, 310

\bibitem[{{Shebalin} {et~al.}(1983){Shebalin}, {Matthaeus}, \&
  {Montgomery}}]{Shebalin1983JPlPh}
{Shebalin}, J.~V., {Matthaeus}, W.~H., \& {Montgomery}, D. 1983, Journal of
  Plasma Physics, 29, 525

\bibitem[{{Smith} {et~al.}(2006){Smith}, {Hamilton}, {Vasquez}, \&
  {Leamon}}]{Smith2006ApJ}
{Smith}, C.~W., {Hamilton}, K., {Vasquez}, B.~J., \& {Leamon}, R.~J. 2006,
  \apjl, 645, L85

\bibitem[{{TenBarge} {et~al.}(2012){TenBarge}, {Podesta}, {Klein}, \&
  {Howes}}]{TenBarge2012ApJ}
{TenBarge}, J.~M., {Podesta}, J.~J., {Klein}, K.~G., \& {Howes}, G.~G. 2012,
  \apj, 753, 107

\bibitem[{{Tessein} {et~al.}(2009){Tessein}, {Smith}, {MacBride}, {Matthaeus},
  {Forman}, \& {Borovsky}}]{Tessein2009ApJ}
{Tessein}, J.~A., {Smith}, C.~W., {MacBride}, B.~T., {Matthaeus}, W.~H.,
  {Forman}, M.~A., \& {Borovsky}, J.~E. 2009, \apj, 692, 684

\bibitem[{{Tu} \& {Marsch}(1995)}]{Tu1995SSRv}
{Tu}, C., \& {Marsch}, E. 1995, Space Science Reviews, 73, 1

\bibitem[{{Tu}(1988)}]{Tu1988JGR}
{Tu}, C.-Y. 1988, \jgr, 93, 7

\bibitem[{{Tu} \& {Marsch}(1990)}]{Tu1990JPlPh}
{Tu}, C.-Y., \& {Marsch}, E. 1990, Journal of Plasma Physics, 44, 103

\bibitem[{{Tu} \& {Marsch}(1993)}]{Tu1993JGR}
---. 1993, \jgr, 98, 1257

\bibitem[{{Tu} {et~al.}(1984){Tu}, {Pu}, \& {Wei}}]{Tu1984JGR}
{Tu}, C.-Y., {Pu}, Z.-Y., \& {Wei}, F.-S. 1984, \jgr, 89, 9695

\bibitem[{{Voitenko} \& {de Keyser}(2011)}]{Voitenko2011NPGeo}
{Voitenko}, Y., \& {de Keyser}, J. 2011, Nonlinear Processes in Geophysics, 18,
  587

\bibitem[{{Whang}(1973)}]{Whang1973JGR}
{Whang}, Y.~C. 1973, \jgr, 78, 7221

\bibitem[{{Wicks} {et~al.}(2010){Wicks}, {Horbury}, {Chen}, \&
  {Schekochihin}}]{Wicks2010MNRAS}
{Wicks}, R.~T., {Horbury}, T.~S., {Chen}, C.~H.~K., \& {Schekochihin}, A.~A.
  2010, \mnras, 407, L31

\bibitem[{{Wicks} {et~al.}(2011){Wicks}, {Horbury}, {Chen}, \&
  {Schekochihin}}]{Wicks2011PhRvL}
---. 2011, Physical Review Letters, 106, 045001

\bibitem[{{Wu} \& {Chao}(2004)}]{Wu2004NPGeo}
{Wu}, D.~J., \& {Chao}, J.~K. 2004, Nonlinear Processes in Geophysics, 11, 631

\bibitem[{{Xiong} \& {Li}(2012)}]{Xiong2012SoPh}
{Xiong}, M., \& {Li}, X. 2012, \solphys, 279, 231

\bibitem[{{Zhao} {et~al.}(2011){Zhao}, {Wu}, \& {Lu}}]{Zhao2011PhPl}
{Zhao}, J.~S., {Wu}, D.~J., \& {Lu}, J.~Y. 2011, Physics of Plasmas, 18, 032903

\bibitem[{{Zhou} \& {Matthaeus}(1989)}]{Zhou1989GeoRL}
{Zhou}, Y., \& {Matthaeus}, W.~H. 1989, \grl, 16, 755

\end{thebibliography}

\end{document}